\documentclass[12pt]{iopart}

\usepackage{iopams}
\usepackage{graphicx}

\begin{document}

\title[Positioning of the rf potential minimum line of a linear Paul trap]{Positioning of the rf potential minimum line of a linear Paul trap with micrometer precision}

\author{P F Herskind\footnote{Present address: Research Laboratory of Electronics, Massachusetts Institute of Technology, Cambridge, Massachusetts 02139, USA}, A Dantan, M Albert, J P Marler and M Drewsen}

\address{QUANTOP, Danish National Research Foundation Center of
Quantum Optics, Department of Physics, University of Aarhus, DK-8000
Denmark.}
\ead{drewsen@phys.au.dk}
\begin{abstract}
We demonstrate a general technique to achieve a precise radial
displacement of the nodal line of the radiofrequency (rf) field in a
linear Paul trap. The technique relies on selective adjustment of
the load capacitance of the trap electrodes, achieved through the
addition of capacitors to the basic resonant rf-circuit used to
drive the trap. Displacements of up to $\sim100~\mu$m with micrometer
precision are measured using a combination of fluorescence images of
ion Coulomb crystals and coherent coupling of such crystals to a
mode of an optical cavity. The displacements are made without
measurable distortion of the shape or structure of the Coulomb
crystals, as well as without introducing excess heating commonly
associated with the radial displacement of crystals by
adjustment through static potentials. We expect this technique to be
of importance for future developments of microtrap architectures and
ion-based cavity QED.
\end{abstract}

\maketitle


\section{Introduction} \label{sect:Introduction}
Radiofrequency (rf) traps provide a simple trapping scenario for
charged particles~\cite{Ghosh1995} and allow for stable confinement
for long periods of time~\cite{Itano1987}. Such traps have proven to be
versatile tools for a wealth of investigations, including quantum
information science~\cite{Blatt_Wineland}, frequency
standards~\cite{Rosenband2008} and cold molecular ion physics~\cite{Staanum2008}.

Due to the time-varying potential inherent in rf traps the trapped
ions undergo rapid motion at the applied rf-frequency. This
so-called micromotion results in Doppler shifts and broadening of
the atomic transitions~\cite{Berkeland1998}, as well as heating of
the ions~\cite{Blumel1989}. Depending on the particular type of trap
used, the rf-field may have either a nodal point or a nodal line for
which the micromotion vanishes. To avoid excess micromotion and
rf-heating of the ions, experiments are often restricted to operate
under conditions where the ions reside in these regions, which are
typically defined by the trap geometry. This may impose severe demands on the manufacturing and
assembly of the trap, should there be a need for the ions to line up
with other components incorporated into the trap structure. In
microtrap architectures~\cite{Stick2006,Seidelin2006} for instance,
it might be desirable to integrate optical fibers for cooling,
manipulation and detection of the ions~\cite{Kim2009}. In the field
of cavity QED with trapped
ions~\cite{Guthohrlein2001,Mundt2002,Herskind2009}, to which this
work applies, the ions should be located within the mode volume of
an optical resonator, which might only be of the order of
$\sim10~\mu$m. In both cases a method for displacing the ions with
$\sim~\mu$m precision over tens of $\mu$m without inducing an excess
of micromotion is therefore of importance to future developments.

One obvious way to achieve this is by physically moving the trap electrodes
and, hence, changing the geometric center of the trap, relative to the
external objects such as the optical fibers or an optical resonator.
In terms of practicality this is, however, not always a viable
approach, especially if these objects are integrated into the trap
structure. Here, we present a general method for translating the
nodal line of a linear Paul trap that does not require any physical
translation of any parts of the trap structure and which can be
accomplished simply through external adjustments of the rf-circuit
residing outside the vacuum chamber. We show that radial
displacements of up to $\pm 100~\mu$m can be achieved without
compromising the trapping of large ion Coulomb crystals and that
positioning of the potential minimum with respect to the axis of an
optical cavity is possible with micrometer precision.

This paper is organized as follows: In section~\ref{sect:setup} we
review the description of ions in a linear Paul trap. In
section~\ref{sect:theory} we present our scheme for displacement of
the nodal line of the rf-field. In section~\ref{sect:trapsetup} we describe the
trap and the technique for measuring the rf-field nodal line position. Section~\ref{sect:results} presents
results based on two different realizations of the scheme, which are
then evaluated and compared against an idealized scenario. In
section~\ref{sect:trap_calibration} we perform a characterization of
our trap and compare the non-displaced configuration with the
displaced one. In section~\ref{sec:tomography} we confirm that the
overlap between the potential minimum and the cavity axis is optimal
within one micrometer by measuring the coherent coupling strength of
prolate ion Coulomb crystals with the cavity field. Finally in
section~\ref{sect:conclusion} we conclude.


\section{The linear Paul trap} \label{sect:setup}
\begin{figure}[htb]
\centerline{\includegraphics[width=1\columnwidth]{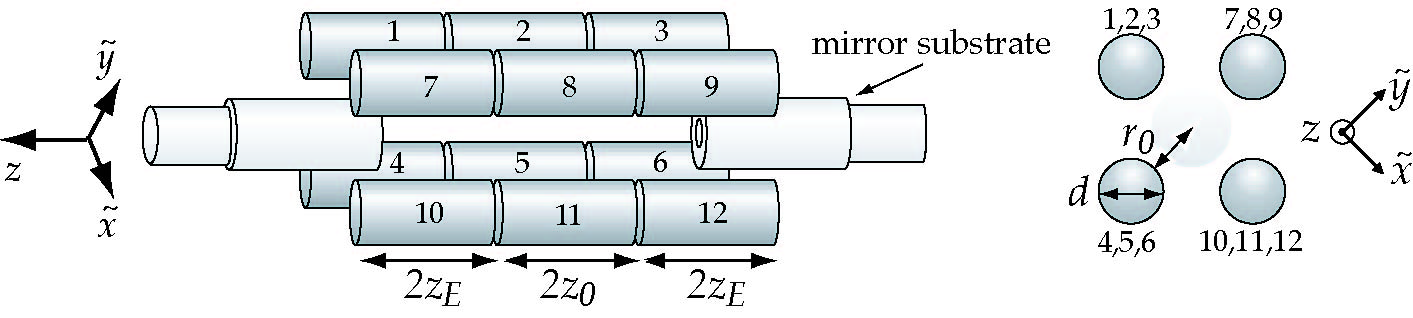}}
\caption{Sketch of the linear Paul trap incorporating an optical resonator. Further details about the
trap can be found in \cite{Herskind2008}. } \label{fig:setup}
\end{figure}
Figure~\ref{fig:setup} shows a schematic of the linear Paul trap used
in the experiments. A special feature of this trap is the
integration of an optical cavity into the trap structure with the
mirrors located in-between the trap electrodes. This trap has been
described in detail elsewhere~\cite{Herskind2008} and here, only a
brief review is presented as reference for the remainder of the
paper. The trap is operated by applying time-varying voltages
$\frac{1}{2}U_{rf}\mathrm{cos}(\Omega_{rf}t)$ to electrodes 1,2,3
and 10,11,12 and $-\frac{1}{2}U_{rf}\mathrm{cos}(\Omega_{rf}t)$
to electrodes 4,5,6 and 7,8,9, where $U_{rf}$ is the amplitude of
the rf-voltage and $\Omega_{rf}$ is the rf-frequency. This gives
rise to a potential in the radial $(\tilde{x}\tilde{y})$-plane of the form
\begin{equation} \label{eq: electric trap pot rf}
\phi_{rf}(\tilde{x},\tilde{y},t)=-\frac{1}{2}U_{rf}\cos(\Omega_{rf}t)\frac{\tilde{x}^2-\tilde{y}^2}{r_0^2},
\end{equation}
where $r_0$ is the inter-electrode inscribed radius. The sectioning of each of the electrode rods
allows for application of a static voltage $U_{end}$ to the
end-electrodes, which provides confinement along the $z$-axis. The
electric potential near the center of the trap is then well
described by
\begin{equation} \label{eq: electric trap pot end}
\phi_{end}(\tilde{x},\tilde{y},z)=\frac{\eta U_{end}}{z_0^2}\left(z^2-\frac{\tilde{x}^2+\tilde{y}^2}{2}\right),
\end{equation}
where $\eta$ is a constant related to the trap geometry and $2z_0$
is the length of the center electrodes (2,5,8,11). From the combined potentials
of (\ref{eq: electric trap pot rf}) and (\ref{eq: electric trap pot
end}) the resulting motion of a single ion near the center of the
trap is described by a Mathieu equation:
\begin{equation} \label{eq:Mathieu_eq}
\frac{\partial^2u}{\partial
\tau^2}+\left[a_u-2q_u\cos(2\tau)\right]u=0,
\hspace{0.5cm}u=\tilde{x},\tilde{y},z.
\end{equation}
We have introduced the following dimensionless parameters:
\begin{eqnarray} \label{eq:tau,a,q}
\tau&=&\frac{\Omega_{rf}t}{2},\hspace{0.5cm} a_{\tilde{x}}=a_{\tilde{y}}=-\frac{a_z}{2}=-4\frac{\eta QU_{end}}{Mz_0^2\Omega_{rf}^2},\nonumber\\
q_{\tilde{x}}&=&-q_{\tilde{y}}=2\frac{QU_{rf}}{Mr_0^2\Omega_{rf}^2},\hspace{0.5cm} q_z=0,
\end{eqnarray}
where Q and M are the charge and mass of the ion, respectively. The
solution to (\ref{eq:Mathieu_eq}) can be written as a Fourier
series, from which regions of stable motion can be
identified~\cite{Drewsen2000}. In general, the trap is operated such that
$\left|a\right|,\left|q\right|\ll1$, in which case the solution to
(\ref{eq:Mathieu_eq}) takes the simple form:
\begin{equation}
  \label{eq:pseudo_eq1}
  u(t) = u_0\cos\left(\omega_u t\right) \left[1-\frac{q_u}{2}\cos\left(\Omega_{rf} t\right)\right],
\end{equation}
where the secular frequency
\begin{equation}
  \label{eq:pseudo_eq2}
  \omega_u = \frac{ \sqrt{q_u^2/2+a_u}}{2}\Omega_{rf}
\end{equation}
has been introduced. From (\ref{eq:pseudo_eq1}), two distinct
types of motion can be identified: the slow \textit{secular} motion at
frequency $\omega_u$ and with amplitude $u_0$, and the superimposed fast
\textit{micromotion} at the rf-frequency $\Omega_{rf}$, which has a
much smaller amplitude due to the smallness of the $q$-parameter.
The amplitude of the secular motion is determined by the thermal
energy of the ion, which can be minimized, e.g., by Doppler laser
cooling.

A constant dc-voltage may also be added to the electrodes in order
to create a static electric field in the radial direction. For
instance, if applied to electrodes 1,2,3, such a field will shift
the location of the potential minimum along $\tilde{x}$. This
modifies the equation of motion as the ion is now displaced from the
nodal line of the rf-field by an amount $u_{dc}$ into a region of
larger micromotion amplitude. (\ref{eq:pseudo_eq1}) thus becomes
\begin{equation}
  \label{eq:pseudo_eq3}
  u(t) = \bigg(u_{dc}+u_0\cos\left(\omega_u t\right)\bigg) \left[1-\frac{q_u}{2}\cos\left(\Omega_{rf} t\right)\right],
\end{equation}
This means that even for the situation where the secular motion is essentially non-present,
the amplitude of the micromotion will still be $\frac{1}{2}u_{dc}q_u$
which may become substantial though the displacement $u_{dc}$ is only of the order of a few microns. This
effect, commonly referred to as excess
micromotion~\cite{Berkeland1998}, shows why radial displacement of
the ion by application of a static potential is undesirable and
motivates an approach based on modification of the potential created
by the rf-field.

For large ion clouds, there will always be some ions with
equilibrium positions in regions not coinciding with the nodal line
of the rf-field and, hence, with large micromotion amplitudes. The
transfer of energy associated with their driven motion into random
thermal motion, through collisions in the crystal, gives rise to an
effective heating rate which has a complex dependence on the
trapping parameters, the temperature and the number of
ions~\cite{Schiffer2000,HerskindThesis}. This counteracts the
Doppler cooling effect and results in temperatures of the ion cloud
that are typically a few tens of mK above the Doppler
limit~\cite{Herskind2009,Drewsen1998,Roth2005}. Nevertheless, under
good cooling conditions the ions may crystallize to form ion Coulomb
crystals and may even exhibit long-range ordered
structures~\cite{Mortensen2006}, expected at temperatures around
10~mK for typical trapping parameters~\cite{Pollock1973}.

In less-ideal scenarios, where the equilibrium position of the ion
cloud does not coincide with the nodal line of the rf-field, the
rf-heating rate can be substantial resulting in
inhomogeneous broadening of the atomic transitions and even failure
to crystallize due to the high temperature of the cloud. Moreover,
this method cannot be used when working simultaneously with
different ion species or isotopes, since the application of a static
field fails to maintain the structural symmetry of multi-component
crystals \cite{Hornekaer2001,Mortensen2007}.

The present paper is focused on general aspects of large ion plasmas and the ability to cool
them into ion Coulomb crystals. The persistent structures of these crystals will serve as one figure of merit for the quality of the displacement of the rf nodal line.
A specific motivation for this work is cavity QED experiments, where the need for maximizing the overlap between a Coulomb crystal and the cavity modevolume is paramount~\cite{Herskind2009}.


\section{Moving scheme} \label{sect:theory}
\begin{figure}[htb]
\centerline{\includegraphics[width=0.6\columnwidth]{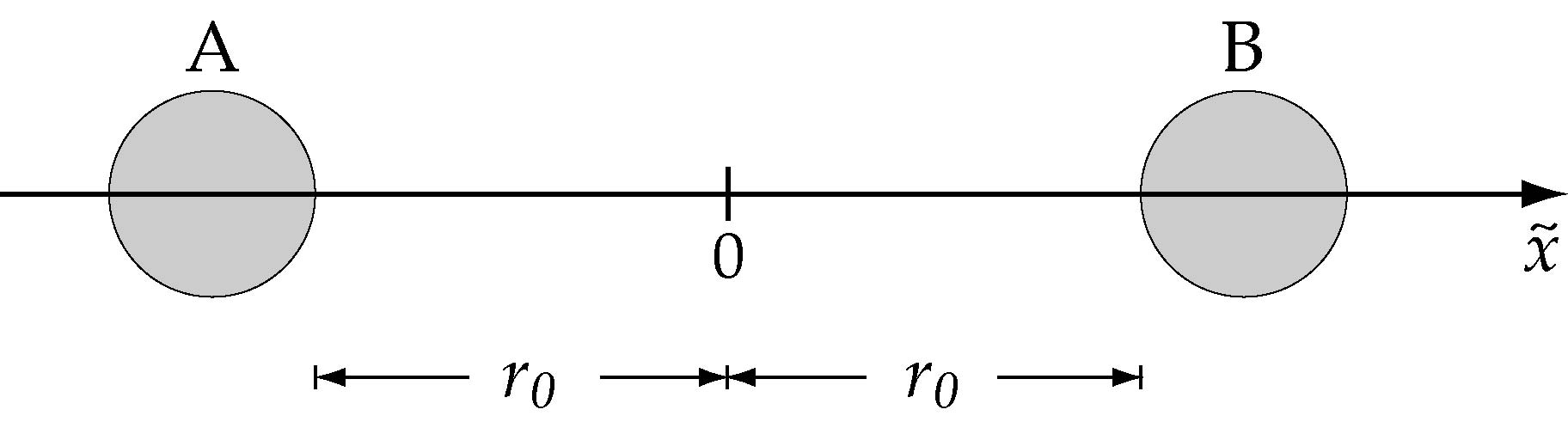}}
\caption{Two electrodes used in the derivation of the location of
the shifted potential minimum. See text for details.}
\label{fig:moving_rf_min}
\end{figure}

For simplicity, we consider the one-dimensional case depicted in
figure~\ref{fig:moving_rf_min}. In the following we shall only analyze
the effect of the rf-fields and assume that no offset with respect
to the nodal line of the rf-potential has been induced by static
fields ($u_{dc}=0$). Instead, we assume that the rf-amplitude on two
electrodes, A and B, can differ by some attenuation factor
$\delta<1$, such that $U_{rf}^B=\delta U_{rf}^A$. The zero-point on
the $\tilde{x}$-axis in figure~\ref{fig:moving_rf_min} indicates the
location of the geometric center at the distance $r_0$ from both
electrodes. The potential from both electrodes falls off with the
inverse of the distance to the electrode and
assuming that the displacement is small compared to the trap
dimensions ($\left|\tilde{x}\right|\ll r_0$), the total potential at $\tilde{x}$ may
be written as
\begin{eqnarray}
\phi(\tilde{x})\varpropto
\frac{U_{rf}^A+U_{rf}^B}{r_0}-\frac{U_{rf}^A-U_{rf}^B}{r_0}\frac{\tilde{x}}{r_0}+\frac{U_{rf}^A+U_{rf}^B}{r_0}\frac{\tilde{x}^2}{r
_0^2}+\Or\Big(\frac{\tilde{x}^3}{r_0^3}\Big). \label{eq:moving_rfmin1}
\end{eqnarray}
Omitting contributions from terms higher than second order, the
location of the minimum $x_0$ is
\begin{eqnarray}
x_0=\frac{U_{rf}^A-U_{rf}^B}{U_{rf}^A+U_{rf}^B}\frac{r_0}{2}=\frac{1-\delta}{1+\delta}\frac{r_0}{2}.
\label{eq:moving_rfmin2}
\end{eqnarray}
For small attenuations ($\delta\approx 1$) one therefore expects the
displacement to be linear, $x_0=(1-\delta)r_0$, and the
potential to remain harmonic around the new minimum. This simple
analysis shows that one can lower the amplitude of the rf-voltage on
the electrode rod in the direction where we wish to move the
potential minimum. The fact that the trap potential retains its
harmonic shape (for small displacements) means that the results of
section~\ref{sect:setup} still hold and that the trap should operate
normally in the new configuration.
\begin{figure}[htb]
\centerline{\includegraphics[width=0.6\columnwidth]{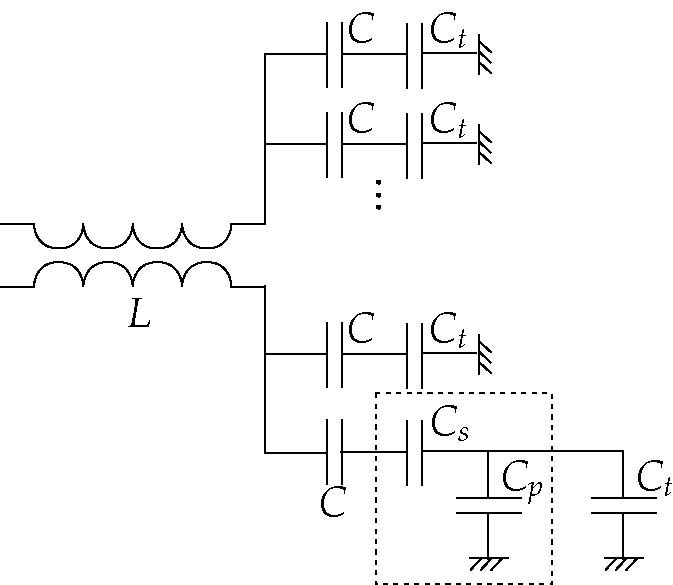}}
\caption{Schematic of the rf resonant circuit. Each electrode is
represented by a capacitance $C_t$. Adjustment of the load on each
electrode can be achieved by adding series and parallel capacitances
$C_s$ and $C_p$ (see dashed box). This will be treated in detail in section~\ref{sect:results}.} \label{fig:LCcircuit}
\end{figure}

The rf-voltage applied to the trap electrodes in our experiments is supplied by a frequency generator (HP 33120A)
and is amplified by an rf-amplifier (Amplifier Research 4W1000) before being transferred to the trap
electrodes via a resonant circuit. In this circuit, the trap itself acts as the capacitative part
of an LRC-circuit which is inductively coupled to the rf-power supply through a ferrite toroid
transformer with a single turn on the source side and 10 turns on the trap side. A diagram of this circuit for the transfer of the rf-voltage to each of the trap
electrodes is shown in figure~\ref{fig:LCcircuit}. It consists of two separate circuits with opposite
phases, which are created from a single rf-input by winding the output coil wires around the ferrite toroid
transformer in opposite directions. Each phase of the rf-voltage is then coupled to a set of six
trap electrodes through capacitors $C$ in series. Missing in the figure, is the part of the circuit used to
add the dc-voltage to the electrodes, omitted for the sake of simplicity. The electric circuit for
each of the electrodes thus consists of a capacitor $C=2.2$~nF and the trap electrode $C_t$ which,
including wires, is typically around 40~pF. The circuit therefore acts as a basic voltage divider
and the voltage on one electrode, $U_e$, with respect to the input voltage, $U_{in}$, is given by
\begin{equation}
U_e=\frac{U_{in}}{1+\frac{C_t}{C}}.
    \label{eq:single_electrode_gain}
\end{equation}
For our parameters, where $C\gg C_t$, the gain across the voltage divider is close to unity
regardless of the exact value of the trap capacitance $C_t$, which may vary slightly from one
electrode to another. Equation (\ref{eq:single_electrode_gain}) shows that lowering the voltage on
selected electrodes can be achieved either by increasing $C_t$ or by decreasing $C$. In section~\ref{sect:results} we will describe how this is done experimentally in both cases. To move the potential
minimum radially, we will make identical changes to the capacitative loads on all three electrodes
of a given electrode rod (e.g. electrodes 1,2,3 in figure~\ref{fig:setup}).


\section{Experimental setup}\label{sect:trapsetup}

\subsection{Linear Paul trap description}
\begin{figure}[htb]
  \centerline{\includegraphics[width=0.6\columnwidth]{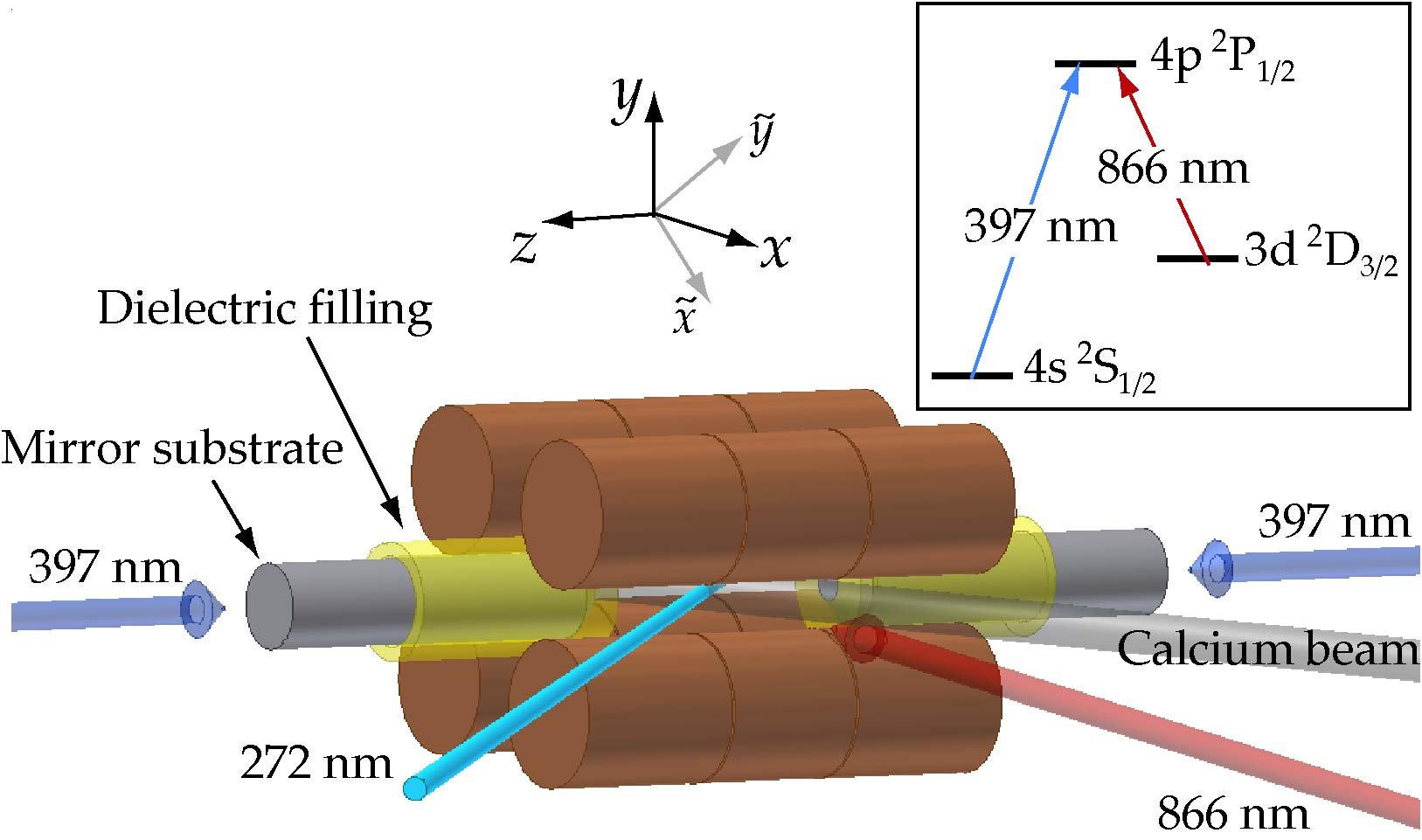}}
\caption{Schematic of the linear Paul trap used in the experiments, along with the beams
used for loading and Doppler cooling of the ions. The inset shows the relevant energy levels of the  calcium ion for Doppler laser cooling.} \label{fig:trapsetup}
\end{figure}
The linear Paul trap used in the experiments is shown in figure~\ref{fig:trapsetup} and has been described in \cite{Herskind2008}. It consists of four segmented cylindrical rods in a quadrupole configuration. The length of the center electrode is
$2z_0=5.0$~mm, and the length of the end-electrodes are $2z_E=5.9$~mm. The electrode diameter is $d=5.2$~mm and the distance from the
trap center to the electrodes is $r_0=2.35$~mm (c.f. figure~\ref{fig:setup}). The trap is operated at a frequency $\Omega_{rf}=2\pi\times4.0$~MHz and the end- and rf-voltages are typically within the range $U_{end}=1-10$~V and $U_{rf}=100-400$~V, respectively. This corresponds to axial and radial secular frequencies in the range $\omega_z=80-260$~kHz and $\omega_r=70-780$~kHz, respectively, and to ion crystal densities between $6.8\times 10^7$ and $1.1\times 10^9$~cm$^{-3}$. The trap also incorporates a moderately high-finesse cavity ($\mathcal{F}\sim 3000$) in between the electrodes, designed to operate on the
3d~$^2$D$_{3/2}$$\rightarrow$4p~$^2$P$_{1/2}$ transition of
Ca$^+$ at 866 nm (see insert of figure~\ref{fig:trapsetup}).

The trap is loaded with $^{40}\mathrm{Ca}^+$ and $^{44}\mathrm{Ca}^+$ ions by isotope selective two-photon photoionization~\cite{Mortensen2004},
by intersecting an atomic beam produced by
an effusive oven with a 272 nm beam at the center of the trap. The $^{40}\mathrm{Ca}^+$ ions produced are subsequently
Doppler-cooled on the
4s~$^2$S$_{1/2}$$\rightarrow$4p~$^2$P$_{1/2}$ transition by two counter-propagating
beams at 397 nm along the
$z$-axis, while in the radial
$(xy)$-plane the ions are sympathetically
cooled through the Coulomb interaction. An 866 nm beam, propagating along the $x$-axis and resonant
with the 3d~$^2$D$_{3/2}$$\rightarrow$4p~$^2$P$_{1/2}$ transition,
is applied to prevent the ions from being shelved into the
metastable D$_{3/2}$ state. Detection of the ions is performed by
imaging spontaneously emitted light at 397 nm onto two image
intensified CCD cameras, monitoring the ions in the $(xz)$- and $(yz)$-planes, respectively.

\subsection{Measurement scheme with fluorescence images}
\begin{figure}[htb]
  \centerline{\includegraphics[width=0.6\columnwidth]{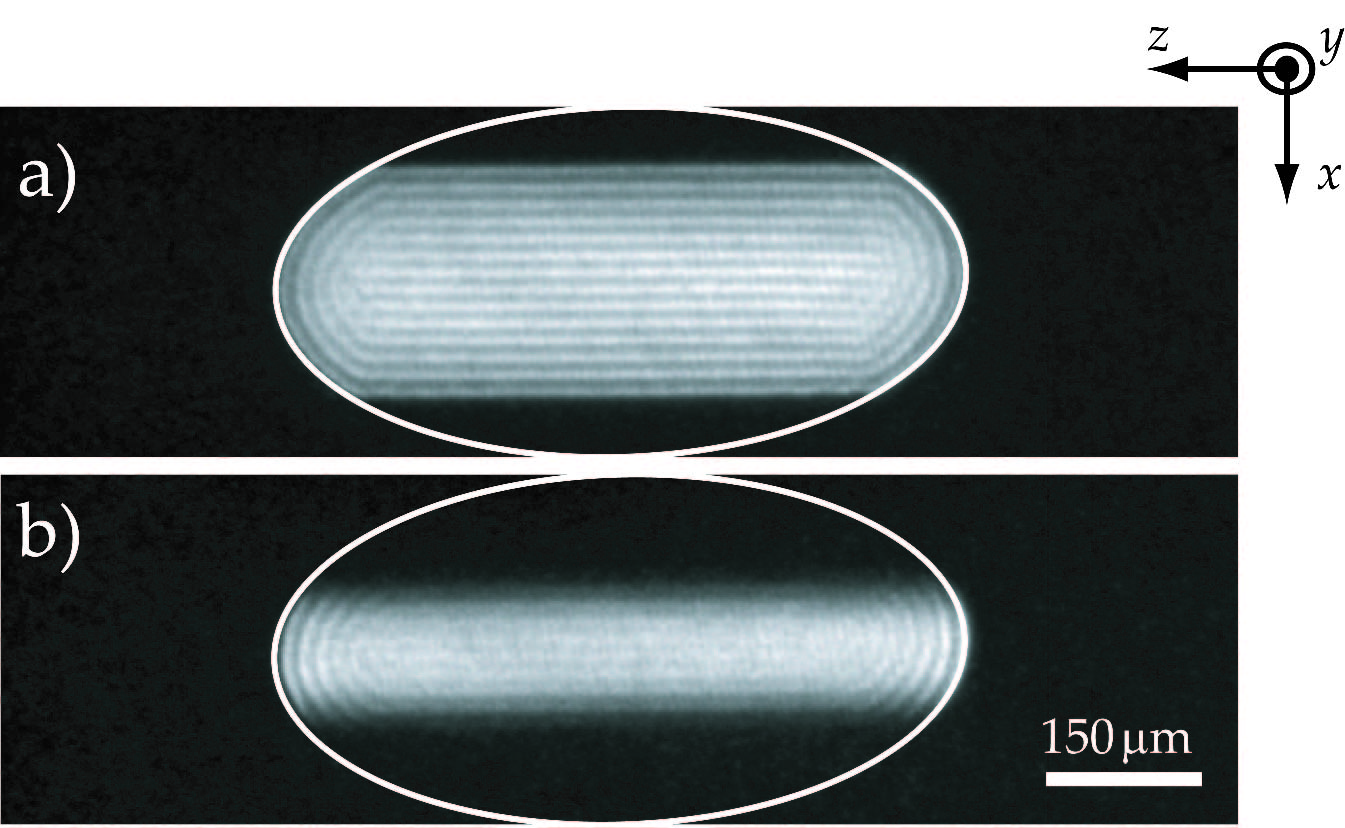}}
\caption{Images of a two-component crystal of $^{40}\mathrm{Ca}^+$
(cooled and visible) and $^{44}\mathrm{Ca}^+$ (outer part, not visible) ions.
From image a) to b) the 866~nm repumper is shifted from the beam
illuminating the entire crystal from the side (along $x$) to the
cavity mode (along $z$). The ellipse indicates the outer boundary of the whole $^{40}\mathrm{Ca}^+$+$^{44}\mathrm{Ca}^+$ crystal.
Exposure time: 1 s.} \label{fig:bicrystal}
\end{figure}

We now turn to the measurement of the rf potential minimum line using fluorescence images of the crystals.
Because of the isotope selectivity of the photoionization scheme used, it is possible
to load simultaneously various isotopes of calcium into the trap~\cite{Herskind2008,Mortensen2004}. The use of a two-component crystal allows for adjusting the static radial trap
potential to coincide with the potential created by the rf-field.
Indeed, since the radial secular trap frequency (\ref{eq:pseudo_eq2})
depends inversely on the mass of the ion, heavier ions are confined
less tightly in the radial plane. As a result, they appear on the
outside of the inner $^{40}\mathrm{Ca}^+$ component. However, this
only happens symmetrically if the static potential created by the
dc-voltages exactly coincides with the rf nodal line. An example of
this is seen in figure~\ref{fig:bicrystal}a), which shows a
two-component crystal consisting of $^{40}\mathrm{Ca}^+$ and
$^{44}\mathrm{Ca}^+$. Only $^{40}\mathrm{Ca}^+$ is being Doppler
laser cooled and visible. The $^{44}\mathrm{Ca}^+$ ions are
sympathetically cooled by $^{40}\mathrm{Ca}^+$ ions. The appearance
of symmetrical dark regions around the $^{40}\mathrm{Ca}^+$ ions is
a signature that the equilibrium location of the ion Coulomb crystal
coincides with the nodal line of the rf-field.

Once the dc potential has been adjusted, the repumping light at 866 nm is shifted from the $x$-axis beam to a beam injected into the cavity ($z$-axis), the frequency of which is close to resonance with the 3d~$^2$D$_{3/2}$$\rightarrow$4p~$^2$P$_{1/2}$ transition. When the repumping light is injected only into the cavity, only the $^{40}\mathrm{Ca}^+$ ions which are located within the cavity modevolume
fluoresce, as can be seen from figure~\ref{fig:bicrystal}b). This allows
for the detection of the cavity mode offset with respect to the ion
Coulomb crystal center in the plane considered and, therefore, the absolute positioning of the nodal line of the rf-field with respect to the cavity axis.

\section{Experimental results}\label{sect:results}

\subsection{Addition of a parallel load}
\begin{figure}[htb]
\centerline{\includegraphics[width=0.8\columnwidth]{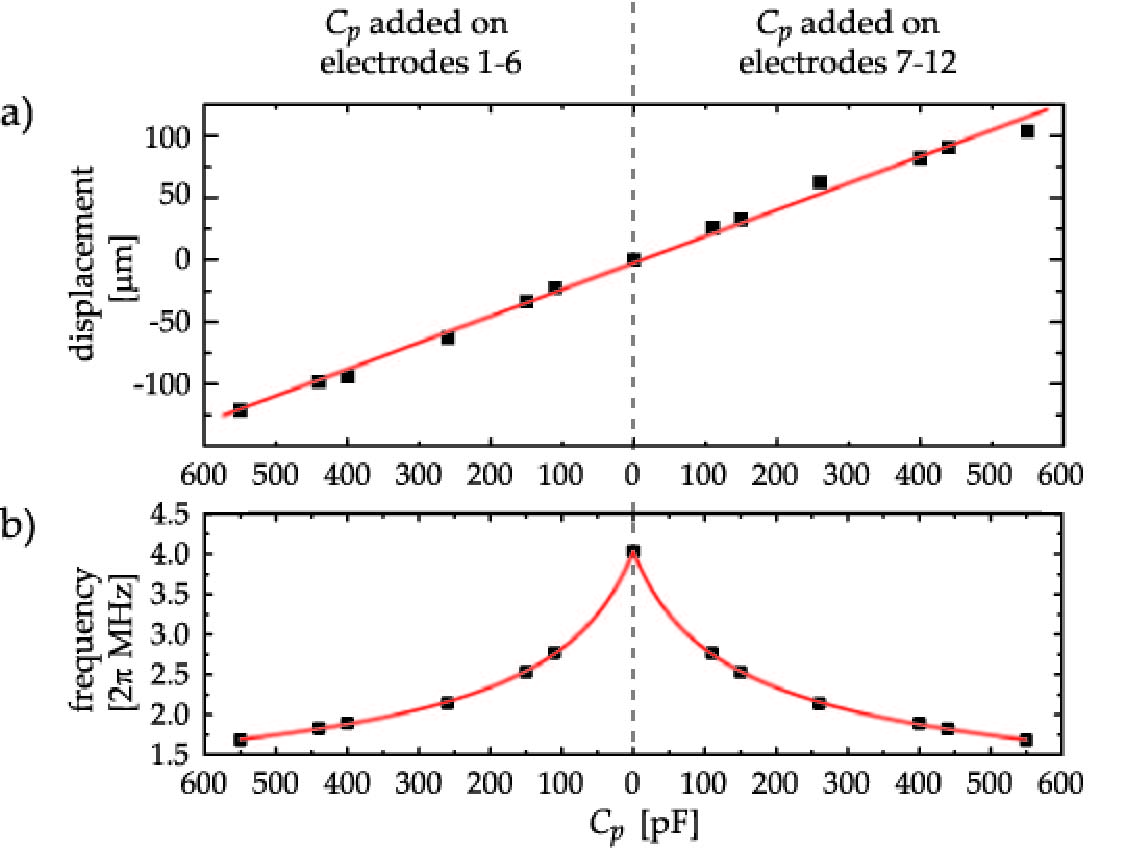}}
\caption{Displacement of the rf-potential minimum found from images of Coulomb crystals by adding a
parallel capacitative load $C_p$ on either electrode 1--6 or
electrodes 7--12. a) Displacement along the $x$-direction versus added load
and linear fit. b) Resonance frequency of the circuit versus added
load. The solid line is of the form $\frac{1}{\sqrt{a+bC_p}}$, where
$a$ and $b$ are free parameters.} \label{fig:moving_parallel_all}
\end{figure}

In this section we will translate the rf potential minimum line by adding loads on the electrodes.
To perform this translation in the $(xy)-$plane and maintain the symmetry of the trap,
the same loads are added to each electrode of a given rod (e.g. 1,2,3). Increasing $C_t$ can be achieved simply by adding a load $C_p$ in parallel, such that
$C_t\rightarrow C_t+C_p$ in (\ref{eq:single_electrode_gain}). The attenuation on each electrode is then given by
\begin{equation}
\frac{U_e'}{U_e}=\frac{1+\frac{C_t}{C}}{1+\frac{C_t+C_p}{C}}\simeq 1-\frac{C_p}{C},
\end{equation}
assuming $C\gg C_t,C_p$. The expected linear scaling of the rf-field nodal line displacement with increasing
value of $C_p$ is confirmed by the results presented in
figure~\ref{fig:moving_parallel_all}a). In these measurements, the
capacitance $C_p$ was added to either electrodes 1--6 or to
electrodes 7--12, as defined in figure~\ref{fig:setup}, thus resulting in a displacement along the $x$-direction by as much as $\mp100~\mu$m. The red line illustrates the linearity of the
displacement.
A drawback of adding a parallel load, however, is that the resonance
frequency of the rf-circuit may be substantially lowered as the load
increases. Figure~\ref{fig:moving_parallel_all}b) shows measurements
of the resonance frequency of the circuit for various values of the added load
$C_p$. As for the displacement, the effect is symmetric with respect
to adding capacitance on either side of the trap. The solid lines
are of the form $\frac{1}{\sqrt{a+bC_p}}$, where $a$ and $b$ are
free parameters. Lowering the resonance frequency may be undesirable in
practice, since, as $\Omega_{rf}$ is lowered for the same values of the stability parameters $q_u$ and $a_u$,
the secular frequency as well as the trapping potential will be lowered, which is often unwanted.

\subsection{Addition of a series load}~\label{sec:series}
As aforementioned, an alternative method consists in adding a
capacitor $C_s$ in series (see figure~\ref{fig:LCcircuit}). In
practice, there is also some coupling to ground associated with
this, which is accounted for by the parallel capacitor $C_p$ ($\sim
10$~pF). The attenuation on one electrode thus becomes
\begin{equation}\label{eq:attenuation_series2}
\frac{U_e'}{U_e}=\frac{1+\frac{C_t}{C}}{1+\frac{(C_t+C_p)(C+C_s)}{CC_s}}\simeq
\frac{1}{1+\frac{C_t+C_p}{C_s}}
\end{equation}
where as previously we have neglected terms according to $C\gg C_t,C_p,C_s$. The resulting
displacement is then inversely proportional to the added series capacitance.

The advantage of this method is that it has a comparatively smaller effect on
the resonance frequency of the circuit. The combined effect of the
added capacitors $C_s$ and $C_p$ is to modify the initial electrode
capacitance as
\begin{equation}
C_{t}\rightarrow C_t'=\frac{C_t+C_p}{1+\frac{C_t+C_p}{C_s}}.
    \label{eq:freq_change_series}
\end{equation}
The LC-circuit resonance frequency for this electrode will then be modified according to
\begin{equation}
\frac{\Omega_{rf}'}{\Omega_{rf}}\sim \sqrt{\frac{C_t}{C_t'}}=\sqrt{\frac{C_t}{C_t+C_p}+\frac{C_t}{C_s}}\label{eq:resonance}
\end{equation}
Provided that $C_s\gg C_t\gg C_p$, the resonance frequency is therefore expected not to be changed significantly.
However, a change, even small, in the resonance frequency for one set of electrodes causes a phase-shift between the
two circuits with opposite phase, which may result in increased micromotion~\cite{Berkeland1998}. (\ref{eq:resonance}) shows that,
with a careful adjustment of the loads $C_s$ and $C_p$, one can keep the resonance frequencies equal for
the two circuits. One can
therefore ensure that there is no phase-shift between the two circuits with opposite phase
after the desired changes in electrode voltages have been obtained.

\begin{figure}[htb]
\centerline{\includegraphics[width=0.6\columnwidth]{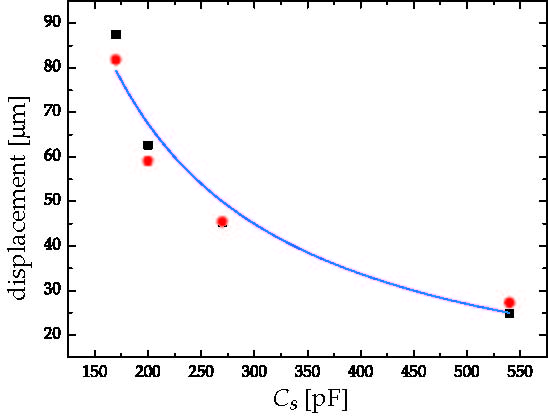}} \caption{Moving the
quadrupole by adding a capacitative load $C_s$ in series. Black squares correspond to the
displacement in the $x$-direction  ($C_s$ added to 1--6) while the red circles are for displacement in the $y$-direction ($C_s$ added to 1--3 and 7--9). The solid line is of the form $\frac{1}{C_s}$.}
\label{fig:moving_series_all}
\end{figure}

Figure~\ref{fig:moving_series_all} shows the resulting displacement of
a single-component Coulomb crystal for various values of $C_s$. Black points correspond to
the displacement in the
$x$-direction when $C_s$ is added
to electrodes 1--6, while the red points are for displacement in the
$y$-direction when $C_s$ is added
to electrodes 1--2--3 and 7--8--9. The solid line is of the form
$\frac{1}{C_s}$, applicable to large values of series capacitance
for which $C_s\gg C_t,C_p$ (c.f. (\ref{eq:attenuation_series2}))
and shows nice qualitative agreement.
\begin{figure}[htb]
  \centerline{\includegraphics[width=0.6\columnwidth]{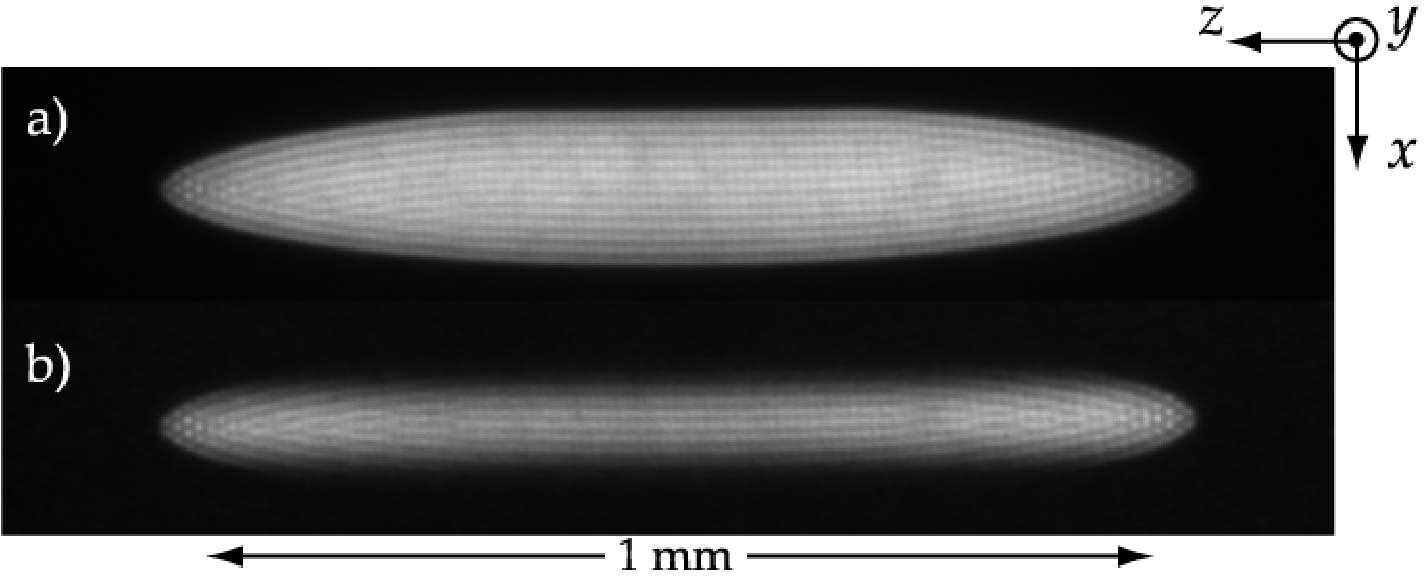}}
\caption{Image of a crystal of $^{40}\mathrm{Ca}^+$ ions in its final position with respect to the cavity mode (adding the set of loads described in section~\ref{sec:series}. From image a) to b) the
866~nm repumper is shifted from the beam propagating along $x$ and illuminating the entire
crystal to the cavity mode along $z$. The rf-field nodal line has been translated by
$\sim 90$ $\mu$m in the $(xz)-$plane, and by $\sim 70$ $\mu$m in the $(yz)-$plane.}
\label{fig:mode_overlap_final}
\end{figure}

Figure~\ref{fig:mode_overlap_final} shows an image where the cavity
mode is clearly seen. From this image, the overlap between the ion
Coulomb crystal and the cavity mode in the $(xz)$-plane is found to be
within $0\pm 2~\mu$m. A second camera system along the
$x$-axis is used for detection of the
overlap in the $(yz)$-plane, which is found to be within $0\pm 8~\mu$m. Because of the optical resolution the images only provide a crude estimate of the
cavity mode position. A more precise method to determine the overlap
between the cavity mode and the crystal center (and therefore the
potential minimum) will be presented in section~\ref{sec:tomography}.

Combination of loads on all electrode rods ((1,2,3),(4,5,6),(7,8,9)
and (10,11,12)) can thus allow for arbitrary positioning of the nodal line of the
rf-field anywhere in the $(xz)$- and $(yz)$-planes. For the crystal in
figure~8 the nodal line has been shifted by
$90~\mu$m along $x$ and $70~\mu$m along $y$, in order to achieve
a near-perfect overlap with the $\mathrm{TEM}_{00}$ mode of the
optical cavity incorporated into the trap. The formation of highly
regular structures indicates that the heating rates and excess
micromotion are still low, which is very encouraging for the
applicability of the method. Furthermore, at these settings, the observed change in the resonance 
frequency, compared to the case without any additional load, is 0.3~$\%$. We estimate the 
phase difference between the rf-fields on electrodes with and without the additional load 
to be less than $1^\circ$ and from the model of reference [6] we estimate the resulting excess 
micromotion to be equivalent to a temperature below 10 mK for typical parameters for 
our trap. Note that for applications where excess micromotion is a critical issue, one 
could e.g. adjust the capacitive loads and use the fluorescence modulation technique of 
[6] to minimize this effect further. 


\section{Trap characterization} \label{sect:trap_calibration}
We now turn to a quantitative characterization of the trap
parameters and compare the results obtained without any additional
load and with the previous set of loads, described in section~\ref{sec:series},  allowing for overlapping the
potential minimum and the axis of the optical cavity integrated in
the trap.

\subsection{Zero-temperature charged liquid model}

Both the radial and the axial frequencies are interesting parameters
to consider when characterizing the trap. However, as our work is
focused on the trapping of large ensembles of ions, we require more
information than just the trap frequencies alone, which are
typically evaluated from motional excitation spectra of a single or
a few trapped ions~\cite{Naegerl1998}. Rather, we seek to confirm
that the assumption of a harmonic potential, as motivated by our
simple model of section~\ref{sect:theory}, is still valid and that
it applies over the entire region in space occupied by the ions. To
this end, we approximate the trapping potential by a harmonic potential with an axial frequency
$\omega_z$ (\ref{eq:tau,a,q}) and an effective radial frequency equal to the radial secular frequency $\omega_r$ given by (\ref{eq:pseudo_eq2}).
Based on this assumption, we model
the ion plasma as a zero-temperature charged liquid
\cite{Turner1987}, which has previously been shown to be an accurate
model for ion Coulomb crystals at temperatures of
$\sim10$~mK~\cite{Hornekaer2001}. According to the model one expects
the following relation between the ratio of the trap frequencies and
the aspect ratio of the crystal, $\alpha\equiv R/L$, where $R$ is
the radius of the crystal and $L$ its axial half-length:
\begin{equation}
\label{eq:trap_freq_ratio_vs_alpha}
        \frac{\omega_z}{\omega_r}= \begin{cases}
{\sqrt{-2\frac{\sinh^{-1}\left(\alpha^{-2}-1\right)^\frac{1}{2}-\alpha\left(\alpha^{-2}-1\right)^\frac{1}{2}}{\sinh^{-1}\left(\alpha^{-2}-1\right)^\frac{1}{2}-\alpha^{-1}\left(\alpha^{-2}-1\right)^\frac{1}{2}}} & \textrm{, for $\alpha<1$}\\
\sqrt{-2\frac{\sin^{-1}\left(1-\alpha^{-2}\right)^\frac{1}{2}-\alpha\left(1-\alpha^{-2}\right)^\frac{1}{2}}{\sin^{-1}\left(1-\alpha^{-2}\right)^\frac{1}{2}-\alpha^{-1}\left(1-\alpha^{-2}\right)^\frac{1}{2}}} & \textrm{, for $\alpha>1$}}
\end{cases}
    \end{equation}
From (\ref{eq:tau,a,q}) and (\ref{eq:pseudo_eq2}) the ratio of
the trap frequencies can be written as
\begin{equation}
    \frac{\omega_z}{\omega_r}=\sqrt{\frac{-(U_{end}-U_{off})}{\beta(\frac{U_{rf}}{2})^2+\frac{1}{2}(U_{end}-U_{off})}},
    \label{eq:trap_freq_ratio_fit}
\end{equation}
where we have introduced $U_{off}$ to account for any static
dc-offsets in the end-voltage $U_{end}$ and where $\beta$ is defined
as
\begin{equation}
  \beta=\frac{q_x^2}{a_x}\frac{U_{end}}{U_{rf}^2}=\frac{-Qz_0^2}{M\eta r_0^4\Omega_{rf}^2},
    \label{eq:beta}
\end{equation}
which for our trap is expected to be
$-(2.29\pm0.06)\times10^{-3}~\mathrm{V}^{-1}$, based on the trap
parameters~\cite{Herskind2008}. 

Equating the right hand sides of (\ref{eq:trap_freq_ratio_vs_alpha}) and (\ref{eq:trap_freq_ratio_fit}) we obtain a relation relating the trap parameters to the aspect ratio of the cold trapped crystals that can be used to test the agreement with the zero temperature charged liquid
model and to calibrate the trap parameters by treating $\beta$ as a free parameter. To achieve this, we trap and cool ion Coulomb crystals of $^{40}\mathrm{Ca}^+$ for a large range of trap parameters ($U_{rf}=100-350$~V and $U_{end}=2-15$~V)  and deduce their aspect ratios $\alpha$ from the recorded images by measuring their radius and length, as described in \cite{Hornekaer2002}. From a fit to the data ($\alpha$ versus $U_{rf}$ and $U_{end}$) we find $\beta=-(2.311\pm0.016)\times10^{-3}~\mathrm{V}^{-1}$, in excellent agreement with our prediction, and $U_{off} = 0.92\pm 0.05~\mathrm{V}$, where the non-zero value can be ascribed to charging effects caused by the UV laser during loading or the trap.

\begin{figure}[htb]
  \centerline{\includegraphics[width=1\columnwidth]{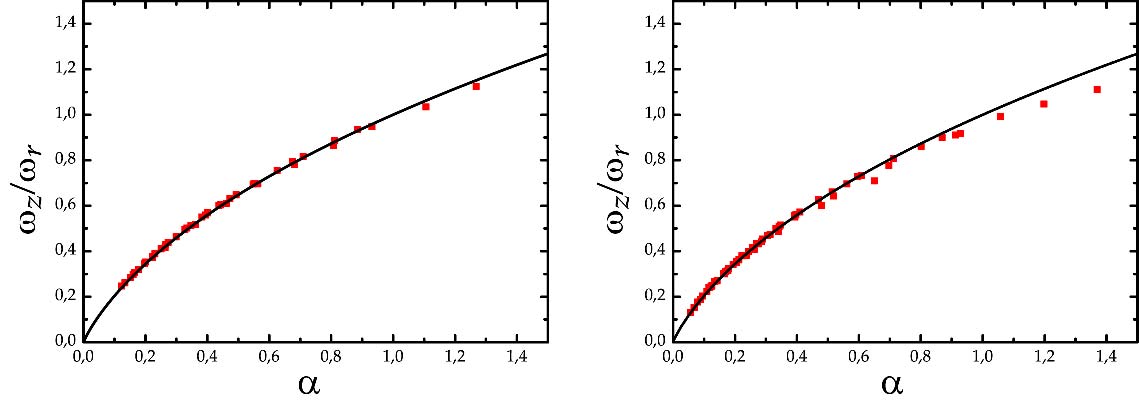}}
\caption{Ratio of trap frequencies versus crystal aspect ratio. a)
For no additional load on the trap electrodes. b) For loads added as described in section~\ref{sec:series}.
The solid black line is the theoretical curve based on the zero
temperature charged liquid model
(\ref{eq:trap_freq_ratio_vs_alpha}) and the red points are
data where $\alpha$ has been measured directly from images of
crystals and $\frac{\omega_z}{\omega_r}$ has been deduced from a fit
to (\ref{eq:trap_freq_ratio_fit}). See text for further details.
The errorbars are within the point size.} \label{fig:trap_freq_exp}
\end{figure}

To give a visual impression of the validity of the fit and the agreement with the zero-temperature charged liquid model, we use the obtained values for $\beta$ and $U_{off}$ to calculate $\frac{\omega_z}{\omega_r}$ via (\ref{eq:trap_freq_ratio_fit}) and, for each measurement, plot this versus $\alpha$ (red, solid squares in figure~\ref{fig:trap_freq_exp}a)). The solid black line represents the theoretical prediction of (\ref{eq:trap_freq_ratio_vs_alpha}) based on the zero temperature charged liquid model and is seen to be in good agreement with the data.

Figure~\ref{fig:trap_freq_exp}b) shows the result of a similar
measurement after the nodal line of the rf-field has been moved
through the addition of serial loads as described in section~\ref{sec:series}. Again, nice
agreement with the zero temperature charged liquid model is seen,
which supports our arguments in the previous section that the trap
potentials should not be distorted by an appreciable amount from
their initial harmonic form. From the fit we find
$U_{off}=(1.20\pm0.03$)~V and
$\beta=-(2.10\pm0.02)\times10^{-3}~\mathrm{V}^{-1}$. Since only the
rf-voltage is modified by our scheme for moving the nodal line, it
is expected that the end-voltage $U_{end}$ and the $a$ parameter are unchanged and that
the ratio $\beta$ is modified only as a result of the attenuation of the
rf-voltage. By comparison with the value found without any
additional load, we deduce an attenuation factor of $0.96\pm0.01$.

\subsection{Wigner-Seitz radius}
\begin{figure}[htb]
\centerline{\includegraphics[width=1\columnwidth]{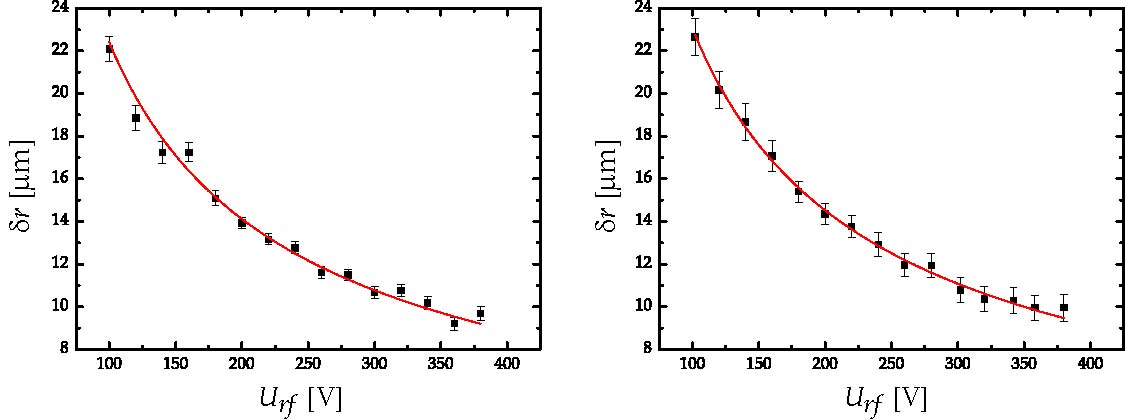}}
\caption{(a) Intershell spacing $\delta r$ measured as a function of
rf-voltage $U_{rf}$ when no additional load is applied to the
electrodes. The red line is a fit based on
(\ref{eq:shellspacing3}) from which the pre-factor is determined
to be $\delta_0=1.484\pm0.010$. (b) $\delta_r$ as a function of $U_{rf}$
when loads are added as described in section~\ref{sec:series}. The red line is a fit based on
(\ref{eq:shellspacing3}) with $\delta_0=1.48$ as a fixed parameter.}
  \label{fig:intershell_measurement}
\end{figure}
For infinitely long crystals of more than three shells, the radial
inter-shell spacing $\delta r$ is predicted by molecular dynamics
(MD) simulations to be constant across the crystal and given by the
relation~\cite{Hasse1990} $\delta r=1.48a_\mathrm{ws}$, where
$a_\mathrm{ws}$ is the Wigner-Seitz radius, defined as
$\frac{4}{3}\pi a_\mathrm{ws}^3=\frac{1}{\rho_0}$.  $\rho_0$ is the
average density of crystal, given by~\cite{Hornekaer2001}
\begin{equation}
    \rho_0=\frac{\epsilon_0U_\mathrm{rf}^2}{Mr_0^4\Omega_\mathrm{rf}^2}= \frac{\epsilon_0\eta}{Qz_0^2}\beta U_\mathrm{rf}^2,
    \label{eq:density}
\end{equation}
from which we can express the inter-shell spacing as
\begin{equation}
\delta r=\delta_0\times\left(\frac{3Qz_0^2}{4\pi\epsilon_0\eta\beta}\right)^{1/3}\times \frac{1}{U_{rf}^{2/3}}.
\label{eq:shellspacing3}
\end{equation}
Here, we have replaced the pre-factor of 1.48 by a fitting parameter
$\delta_0$, which allows us to test the validity of the molecular
dynamics simulations for our trap.

Figure~\ref{fig:intershell_measurement}a) shows the result of
measurements of the inter-shell spacing $\delta r$ for different
values of the rf-voltage in the configuration where no additional
load has been applied to the electrodes. To mimic the notion of
``infinitely long'' in the MD simulations, low aspect ratio crystals
of 1.5-2~mm length were employed for these measurements. $\delta r$
is determined from the recorded crystal images as described in
\cite{Drewsen1998} and the red line shows the result of a fit
to the data based on (\ref{eq:shellspacing3}). The fit gives a
value for the pre-factor of $\delta_0=1.484\pm0.010$, which is a
significant improvement in the determination of this factor over
previous measurements~\cite{Drewsen1998} and a strong support of the
MD simulations of \cite{Hasse1990}.

A similar set of data has been obtained for the configuration where
the serial loads have been added.
Figure~\ref{fig:intershell_measurement}b) shows the results, where for
the fit we inserted $\delta_0=1.48$ and used $\beta$ as a free
parameter. Again, nice agreement with the model is found and a value
of $\beta=-(2.137\pm0.045)\times 10^{-3}~\mathrm{V}^{-1}$ is
obtained in good agreement with the measurements of the previous
section.

\section{Overlap between the cavity mode and the potential
minimum}\label{sec:tomography}
\begin{figure}[htb]
  \centerline{\includegraphics[width=1\columnwidth]{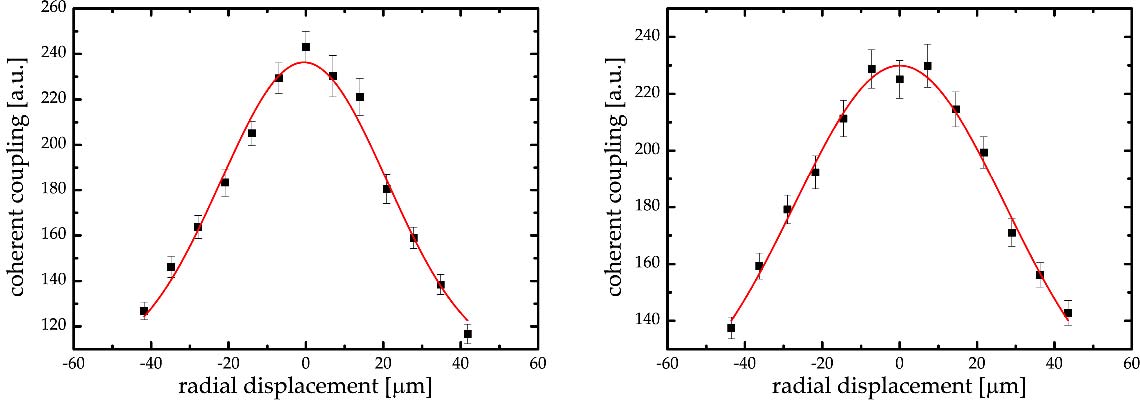}}
\caption{Coherent coupling strength as a function of the crystal
radial displacement from the rf nodal line in the $(xz)$-plane
(left) and the $(yz)$-plane (right). The solid lines are Gaussian
fits, yielding horizontal and vertical offsets of $0.0\pm0.7$ $\mu$m
and $-0.5\pm0.5$ $\mu$m, respectively.} \label{fig:tomography}
\end{figure}
In order to measure more precisely the overlap between the rf
potential minimum and the cavity mode one can make use of the
coherent coupling of a thin, prolate crystal with the cavity
fundamental TEM$_{00}$ field mode, as demonstrated in
\cite{tomography}. By translating radially a crystal whose
radius is smaller than the waist of the cavity and monitoring the
change in the coherent coupling strength with the cavity field one
can reconstruct the transverse mode profile of the cavity field. We
performed similar experiments to those of \cite{tomography} and
translated the ions along the $x$- and $y$-axes by application of a
suitable static electric field to electrodes (1,2,3) and (4,5,6).
The coherent coupling was measured by scanning the cavity length
around atomic resonance and injecting a probe field at the single
photon level, resonant with the 3d$^2$D$_{3/2}\rightarrow4$p$^2$P$_{1/2}$
transition~\cite{Herskind2009}. Due to the absorption induced by the
ions the width of the cavity reflection spectrum is broadened by an
amount proportional to the coherent coupling strength, which depends
on the overlap between the crystal and the cavity
modevolume~\cite{tomography}. Figure~\ref{fig:tomography} shows the
variation of the coherent coupling strength with the radial
translation of the crystal along two orthogonal axes. The positions
of the maxima gives an offset of less than a micrometer of the rf
nodal line ($0.5\pm 0.6$ $\mu$m). This confirms the near-optimal
positioning of the potential minimum with respect to the axis of the optical cavity.

\section{Conclusion} \label{sect:conclusion}
In conclusion we have developed a method for radially translating
the rf nodal line of a linear Paul trap based on selective adjustment
of capacitative loads on the trap electrodes. Two different methods
were analyzed and tested and in both cases the results were
well-accounted for by simple models. By appropriate design of the
resonant circuit for the rf-voltage, all adjustments can be made
outside the vacuum chamber which makes the method attractive from a practical point of view. In the second scheme, adding both parallel and series
loads allowed for an arbitrary translation of the potential minimum
as well as a precise control of the relative phase between the two
rf-circuits and the resonance frequency.

Based on the images of the trapped ion Coulomb crystals, we observed
no additional heating effects, caused by excess micromotion, as one
would expect, had the ion location been shifted through adjustment
of static radial potentials. Furthermore, the validity of both the
zero-temperature charged liquid model and the scaling of the
inter-shell spacing with the Wigner-Seitz radius, when the position
of the ion crystal was shifted, supports the non-invasiveness of the
technique. Incidentally, we have found that this analysis provides a
value for the pre-factor in the relation between the inter-shell
spacing and the Wigner-Seitz radius, with an uncertainty that, to
our knowledge, is the lowest obtained thus far, and which is in
perfect agreement with MD simulations.

Finally, we employed the coherent coupling between the ion Coulomb
crystal and the TEM$_{00}$ mode of the cavity to obtain
high-resolution measurements of the location of the crystal with respect to the cavity mode axis. This made it possible to position the ion
Coulomb crystal in the cavity mode with a precision at the
micrometer level. We believe this technique will become of high
value for cavity QED based ion-photon interfaces and for the
development of microtrap architectures for ion-based quantum
information science.

\ack
The authors would like to thank Erik S\o ndergaard for the
development of the original rf-circuit and for useful discussions on
the modification of the design. We acknowledge financial support
from the Carlsberg Foundation and the Danish Natural Science
Research Council through the ESF EuroQUAM project CMMC.

\end{document}